\documentclass[reprint,aip,jcp]{revtex4-1}

\usepackage{graphics}
\usepackage{graphicx}

\begin{document}

\title{Fragile-to-strong transition in liquid silica}

\author{Julian Geske}
\affiliation{geske@fkp.tu-darmstadt.de}
\address{Institut f\"ur Festk\"orperphysik, Technische Universit\"at 
Darmstadt, Hochschulstr. 6, 64289 Darmstadt, Germany}
\author{Barbara Drossel}
\author{Michael Vogel}
\address{Institut f\"ur Festk\"orperphysik, Technische Universit\"at 
Darmstadt, Hochschulstr. 6, 64289 Darmstadt, Germany}

\begin{abstract}
We investigate anomalies in liquid silica with molecular dynamics simulations and present evidence for a fragile-to-strong transition at around 3100\,K-3300\,K. To this purpose, we studied the structure and dynamical properties of silica over a wide temperature range, finding four indicators of a fragile-to-strong transition. First, there is a density minimum at around 3000\,K and a density maximum at 4700\,K. The turning point is at 3400\,K. Second, the local structure characterized by the tetrahedral order parameter changes dramatically around 3000\,K from a higher-ordered, lower-density phase to a less ordered, higher-density phase. Third, the correlation time $\tau$ changes from an Arrhenius behavior below 3300\,K to a Vogel-Fulcher-Tammann behavior at higher temperatures. Fourth, the Stokes-Einstein relation holds for temperatures below 3000\,K, but is replaced by a fractional relation above this temperature. Furthermore, our data indicate that dynamics become again simple above 5000\,K, with Arrhenius behavior and a classical Stokes-Einstein relation.
\end{abstract}

\maketitle

\section{Introduction}

Network-forming liquids, such as H$_2$O, SiO$_2$, Si, Ge, Sb, Bi, and Ga,\cite{tanaka2002} show complex structural and dynamical features. Their specific properties are due to the fact that they form nearest-neighbor bonds that are strongly directional, with specific values for the bond lengths and bond angles. The number of such bonds is smaller than the usual number of neighbors in a liquid; it has for instance the value of four in tetrahedral networks, as formed by water and silica.\cite{angell2000} Therefore, the density of the liquid decreases as the bond network is formed, and the structure becomes locally more ordered. Since the formation of the network implies  a decrease in entropy and in potential energy, the formation of the low-density networks is favored for lower temperatures and pressures, leading to various anomalies. The most famous of these anomalies, the density maximum of water, has been known for a long time.\cite{fahrenheit1724} 

Water shows additional anomalies, for instance, in the isothermal compressibility\cite{speedy1976} and the heat capacity.\cite{angell1983} They are often particularly pronounced for the supercooled liquid.\cite{mishima1998,angell2000,prielmeier1987} The unusual structural and thermodynamical properties of water are accompanied by anomalous dynamical behavior, such as a diffusion anomaly.\cite{angell2000,prielmeier1987} Moreover, it was proposed that water exhibits a fragile-to-strong (FS) transition in the supercooled regime,\cite{ito1999} which means that the temperature dependence of the structural ($\alpha$) relaxation changes from Vogel-Fulcher-Tammann (VFT) to Arrhenius behavior upon cooling, but an experimental observation of this phenomenon is hampered by a high tendency for crystallization. All these features of water can be explained by using simple models that allow for three nearest-neighbor states with different energy and entropy.\cite{heckmann2012,heckmann2013}

Silica resembles water in several respects. SiO$_2$ and H$_2$O have the same stoichiometry with two A atoms and one B atom. Furthermore, the A-B-A bond angle (O-Si-O or H-O-H) is about the tetrahedral angle which, together with 2:1 stoichiometry, allows for a formation of tetrahedral networks. Finally, like water, silica is polar because of substantial partial charges of the silicon and oxygen atoms. Therefore, one may expect that silica shares many properties with water.\cite{debenedetti1996} Consistently, a density maximum was also seen in computational\cite{poole1997,shell2002} and experimental\cite{brueckner1970,angell1976} studies on silica. It occurs around 1800\,K at 1\,atm.\cite{angell1976} In simulations, also other anomalies were reported, \cite{saikaPhase2004} for example in the specific heat.\cite{poole1997,sharma2006} Yet, as compared to water, silica has a lower tendency for crystallization and, hence, it is a better glass-former.

Silica is usually considered as the paradigm of a strong glass former.\cite{richet1984,angell1991} Over a wide range, the temperature dependence of self-diffusion coefficients and structural relaxation times is described by an Arrhenius law with activation energies of ca.\ 5\,eV. However, experimental\cite{roessler1998, hess1996,sonneville2013} and computational\cite{horbach1999, horbachRelaxation2001, saika2001, saikaFree2004, vogelLetter2004, vogel2004, heuer2004} studies found that the strong behavior changes to a fragile one when silica is heated to temperatures above 3000\,K. Similar to the situation for water, it is currently highly debated whether this FS transition is a consequence of a true phase transition or a smooth crossover. Evidence for a transition from a high-density liquid (HDL) to a low-density liquid (LDL) in silica \cite{saika2000,lascaris2015,sastry2003} supports the idea that the system may exhibit a liquid-liquid phase transition in the viscous regime. In fact, the above-mentioned simple three-state models \cite{heckmann2013} lead generically to a thermodynamic phase transition between the HDL and LDL silica forms. However, with different parameter choices the phase transition is replaced by a crossover, and therefore the question is yet unsettled. For silica, crystallization does not interfere with studies of the liquid state in the relevant temperature range. Thus, experimental studies should, unlike for water, be able to determine whether a liquid-liquid phase transition occurs for silica. The relatively high temperatures of ca.\ 3000\,K are a serious drawback to the application of many experimental techniques, though. Conversely, a FS crossover occurring at high temperatures and fast dynamics, i.e., for structural relaxation times of the order of 1\,ns, provides ideal conditions for molecular dynamics (MD) simulations approaches.\cite{horbach1999, horbachRelaxation2001, saika2001, saikaFree2004, vogelLetter2004, vogel2004, heuer2004}
 
In this paper we extend the existing theoretical studies on silica by performing MD simulations for system sizes and temperature ranges that are considerably larger than before, and by combining the investigation of structural and dynamical properties. Our main finding is a clear change from Arrhenius to VFT behavior for the structural relaxation times as temperature is increased. This change occurs approximately at the same temperature as the density turning point and a change from a low-density silica with high local tetrahedral order to a high-density silica with less local order. For very high temperatures, the structural relaxation reverts again to an Arrhenius law. The change from Arrhenius behavior to VFT and back to Arrhenius is accompanied by a transition from a classical Stokes-Einstein relation to a fractional one and back to a classical one as the temperature increases. 

\section{Simulation Details}\label{sec:simulation_details}

We used the \textit{NAMD} simulation software package. The MD simulations of SiO$_{2}$ in the bulk state were performed with $N=17496$ atoms (5832 silicon and 11664 oxygen atoms) in a cubic volume centered about the origin, with periodic boundary conditions. All simulations were performed in the $NpT$ ensemble. To keep the temperature $T$ and the pressure $p$ constant, the Langevin thermostat  and the Langevin-Piston-Nos\'{e}-Hoover barostat were used. The time step of the integration was set to 1\,fs. For temperatures above 5000\,K the time step had to be adjusted to 0.5\,fs, because otherwise atoms come too close to each other within one integration step. Simulations were performed for up to 135\,ns, depending on temperature. The pressure was set to 1\,bar. The masses of oxygen and silicon atoms were set to 15.9994\,u and 28.0850\,u respectively. Prior to data acquisition the system was equilibrated. At all temperatures, we ensured that the equilibration times exceed the structural relaxation times.

We used a modified version of the interaction potential proposed by van Best, Kramer, and van Santen (BKS).\cite{bks} The BKS potential proved well suited to reproduce physical properties of silica.\cite{horbach1999,horbachRelaxation2001,saika2000} It describes the particle interactions by the combination of a Coulomb term and a Buckingham term $V_{\mathrm{B}}$. The latter is given by 
\begin{equation}
V_{\mathrm{B}}(r)=A_{\mathrm{ij}} \ \exp (-b_{\mathrm{ij}} r)-\frac{c_{\mathrm{ij}}}{r^6} \  \mathrm{.}
\label{eq:buck}
\end{equation}
The use of a modified BKS potential was necessary because the Coulomb and the Buckingham interactions between oxygen and silicon are attractive for small distances. Therefore, we had to ensure that the atoms do not get unphysically close to each other, in particular, at high temperatures where their kinetic energies suffice to cross the energy barrier separating this fully attractive region from the usual minimum region of the Buckingham potential. The modified potential therefore consists of the regular Buckingham potential at larger distances with a continuously differentiable transition to the repulsive part of a Lennard-Jones potential at smaller distances, 
\begin{equation}
V_{\mathrm{LJ}}(r)=\frac{C_{\mathrm{ij}}^{(12)}}{r^{12}} - \frac{C_{\mathrm{ij}}^{(6)}}{r^6} \  \mathrm{.}
\label{eq:lj}
\end{equation}
Specifically, we switch between both potentials at a distance $r_{\mathrm{max}}=1.09\,\mathrm{\AA}$ for silicon-oxygen and at $r_{\mathrm{max}}=1.50\,\mathrm{\AA}$ for oxygen-oxygen interaction, corresponding to the maximum of the Buckingham potential. The parameters used in our simulations are listed in Table~\ref{tab:buck} and Table~\ref{tab:lj}.  

We checked how many atoms penetrated into the region $r<r_{\mathrm{max}}$.  At 10000\,K, we found that 0.50 \% of the  atoms reside in the region where the Buckingham potential was modified. This percentage is relatively small, nonetheless it can change the behavior of the system. However, the percentage decreases fast with lower temperatures, e.g, at 7000\,K, this percentage amounts to only 0.07\%, suggesting that the modification of the potential does not play a role at $T\leq 7000$\,K.
\\
We simulated the following temperatures: between 2300\,K and 3500\,K in 100\,K steps and additionally 3800\,K, 4100\,K, 4400\,K, 4700\,K, 5000\,K, 5500\,K, 6000\,K, 6500\,K, 7000\,K, 8000\,K, 9000\,K, and 10000\,K.

\begin{table*}[ht]
\centering
\begin{tabular}{c|c|c|c|c|||c}
Atom 1 & Atom 2 & $A_{\mathrm{ij}}$ in kcal/mol & $b_{\mathrm{ij}}$ in \AA & $c_{\mathrm{ij}}$ in \AA $^6$ kcal/mol & charge \\ \hline
O & O & \ 32025 & 2.76 & 4036 & $q_{\mathrm{O}}=-1.2 e$\\
Si & O & 415166 & 4.87 & 3079 & $q_{\mathrm{Si}}=2.4 e$\\
Si & Si & 0 & - & 0 & \\
\end{tabular}
\caption{\label{tab:buck} Parameter values for the Buckingham Potential (Eq.~(\ref{eq:buck})) \cite{bks}}
\end{table*}

\begin{table*}[ht]
\centering
\begin{tabular}{c|c|c|c}
Atom 1 & Atom 2 & $C_{\mathrm{ij}}^{(6)}$ in kcal \AA $^6$/mol & $C_{\mathrm{ij}}^{(12)}$ in kcal \AA $^{12}$/mol \\ \hline
Si & O & \ 1124.08 & \ 13776 \\
O & O & -2275.22 & 281743 \\
Si & Si & 0 & 0 \\
\end{tabular}
\caption{\label{tab:lj} Parameter values for the Lennard-Jones Potential in Eq.~(\ref{eq:lj}). \cite{bormuth2012}}
\end{table*}

\section{Studied measures of structure and dynamics}\label{sec:Evaluated_Parameters_and_Functions}

\subsection{Tetrahedral order}

The extent to which the atoms order locally can be quantified by the tetrahedral order parameter. The tetrahedral order $q_{i}$ of the silicon atom $i$ is defined as \cite{chau1998,errington2001}
\begin{equation}
q_{i}=1-\frac{3}{8} \sum_{j=1}^{3} \sum_{k=j+1}^{4} \left (\cos \Theta_{ijk} + \frac{1}{3} \right )^2 \mathrm{ ,}
\end{equation}
where $\Theta_{ijk}$ denotes the angle between the silicon atom $i$ and two of its four nearest silicon atoms $j$ and $k$. Possible values of $q_{i}$ lie between -3 and 1, where the latter value is found for a perfect local tetrahedral structure. The tetrahedral order parameter $Q$ of the whole system is defined as the average over all silicon atoms. If $f(q_i)$ denotes the distribution of the $q_i$ in the system, $Q$ is obtained from the integral
\begin{equation}
Q= \langle q_{i} \rangle =  \int_{-3}^1 f(q_{i})\ q_{i} \ \mathrm{d}q_{i} \, .
\end{equation}
A value of $Q=1$ means that all atoms are arranged in a perfect tetrahedral structure, while $Q=0$ for an ideal gas. Closely related to the tetrahedral order parameter is the tetrahedral entropy. It is defined as\cite{kumar2009} 
\begin{equation}
S_{Q}(T)=S_{0}+\frac{3}{2}\ k_{\mathrm{B}}\ \int_{-3}^{1}\ \ln (1-q_i)\ f(q_i,T)\ \mathrm{d}q_i \mathrm{ .}
\end{equation}
Here, $S_{0}$ is a constant and $k_{\mathrm{B}}$ is the Boltzmann constant.

\subsection{Molecular dynamics}

In order to analyze the structural relaxation of the studied silica melt, we calculate the incoherent intermediate scattering function (ISF). For isotropic systems it can be obtained according to
\begin{equation}
F_{\mathrm{s}}(q,t)=\left\langle \frac{\sin\  \left(q\cdot \left|\vec{r_i}(t_0+t)-\vec{r_i}(t_0) \right| \right ) }{q\cdot \left|\vec{r_i}(t_0+t)-\vec{r_i}(t_0)\right|}  \right\rangle
\label{eq:isf}
\end{equation}
from the particle displacement vectors $\vec{r_i}(t_0+t)-\vec{r_i}(t_0)$ during a time interval $t$. Here, the angular brackets denote the averages over all atoms of a given species and various time origins $t_0$. $F_{\mathrm{s}}(q,t)$ probes particle displacements on a length scale determined by the absolute value of the scattering vector $q$. We use $q=2.0\,$\AA $^{-1}$ corresponding to the nearest neighbor distance between silicon atoms, which is roughly $3.1\,$\AA. We define the structural relaxation time $\tau$ as the time after which the ISF has decayed from 1 to  $1/e$.

For strong glass formers, the structural relaxation time $\tau$ follows an Arrhenius law 
\begin{equation}
\tau=\tau_{0} \cdot \exp \left(\frac{E_\mathrm{A}}{k_{\mathrm{B}}T } \right)\ \mathrm{ .}
\label{eq:arrhenius}
\end{equation}
Here, $E_{\mathrm{A}}$ is the activation energy and $\tau_{0}$ is related to the attempt frequency. In the case of a fragile liquid, the temperature dependence of $\tau$ is often well described by a VFT behavior
\begin{equation}
\tau=\tau_{0} \cdot \exp \left(\frac{E_\mathrm{VFT}}{k_{\mathrm{B}}\cdot (T-T_{\infty}) } \right)\ \mathrm{,}
\label{eq:vft}
\end{equation}
implying a divergence of the structural relaxation time at the temperature $T_{\infty}$.  

The self part of the van Hove correlation function $G_\mathrm{s}(\vec{r},t)$ is well suited to explore the mechanism for the structural relaxation. It is defined as \cite{hansen1986}
\begin{equation}
G_\mathrm{s}(\vec{r},t)=\langle \delta \left[\vec{r}_{i}(t+t_0)-\vec{r}_{i}(t_0)-\vec{r}\,  \right] \rangle \ \mathrm{ .}
\label{eq:gself}
\end{equation}
Thus, for isotropic systems the quantity $4\pi r^2G_\mathrm{s}(r,t)$ measures the probability density that an atom has travelled the distance $r=|\vec{r}|$ during the time interval $t$. 

The fractional Stokes-Einstein relation for supercooled liquids is \cite{fujara1994,roessler1990}
\begin{equation}
D \cdot \left( \frac{\tau}{T} \right)^{\theta} = C  \ \ \mathrm{.}
\label{eq:stokes-einstein}
\end{equation}
Here $D$ is the diffusion coefficient and $C$ a constant. A breakdown of the classical Stokes-Einstein relation occurs if $\theta \neq 1$.\cite{henritzi2015}

\section{Results}\label{sec:results}

\subsection{Structure}

\begin{figure}[ht]
\centering
\includegraphics[scale=0.4]{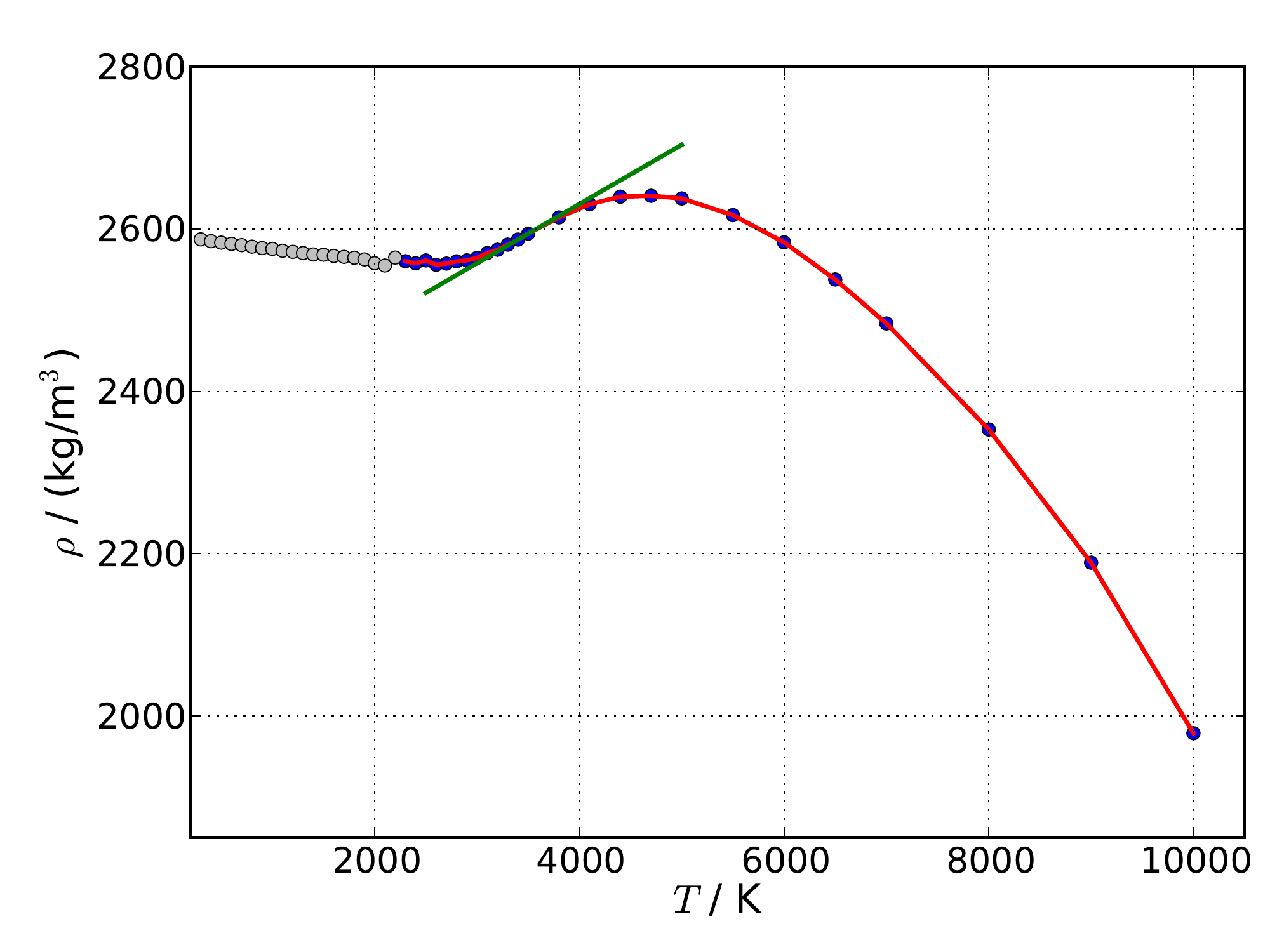}
\caption{\label{fig:density}  Density of liquid silica at a pressure of 1\,bar over a wide temperature range. Remarkable are the density extrema: a minimum at $T \approx 3000$\,K and a maximum at $T\approx4700$\,K. The turning point is at $T \approx 3400$\,K. The tangent for this temperature is displayed as solid line. For all temperatures below 2300\,K the relaxation times are too long to fully equilibrate the system and, hence, the corresponding density values, which are marked in gray, should be taken with caution. }
\end{figure}

The temperature-dependent density of the studied silica melt is shown in Fig.~\ref{fig:density}. In contrast to most other liquids, where the density decreases monotonously with increasing temperature due to the increasing kinetic energy, BKS silica shows a local density minimum at about 3000\,K, followed by a local maximum at approximately 4700\,K. The steepest increase in density occurs at $T\approx 3400$\,K.
The density anomaly of silica melt was studied in both simulations\cite{poole1997,shell2002} and experiments.\cite{angell1976,brueckner1970} An experimental study reported a density maximum at 1800\,K for ambient pressure,\cite{angell1976} which is much lower than our value. This difference illustrates the fact that force fields in MD simulations are always optimized for certain quantities. Obviously, the density maximum of silica was not a reference value for the parametrization of the BKS potential. Nevertheless, this model correctly reproduces the qualitative behavior.
 
\begin{figure}[ht]
\centering
\includegraphics[scale=0.4]{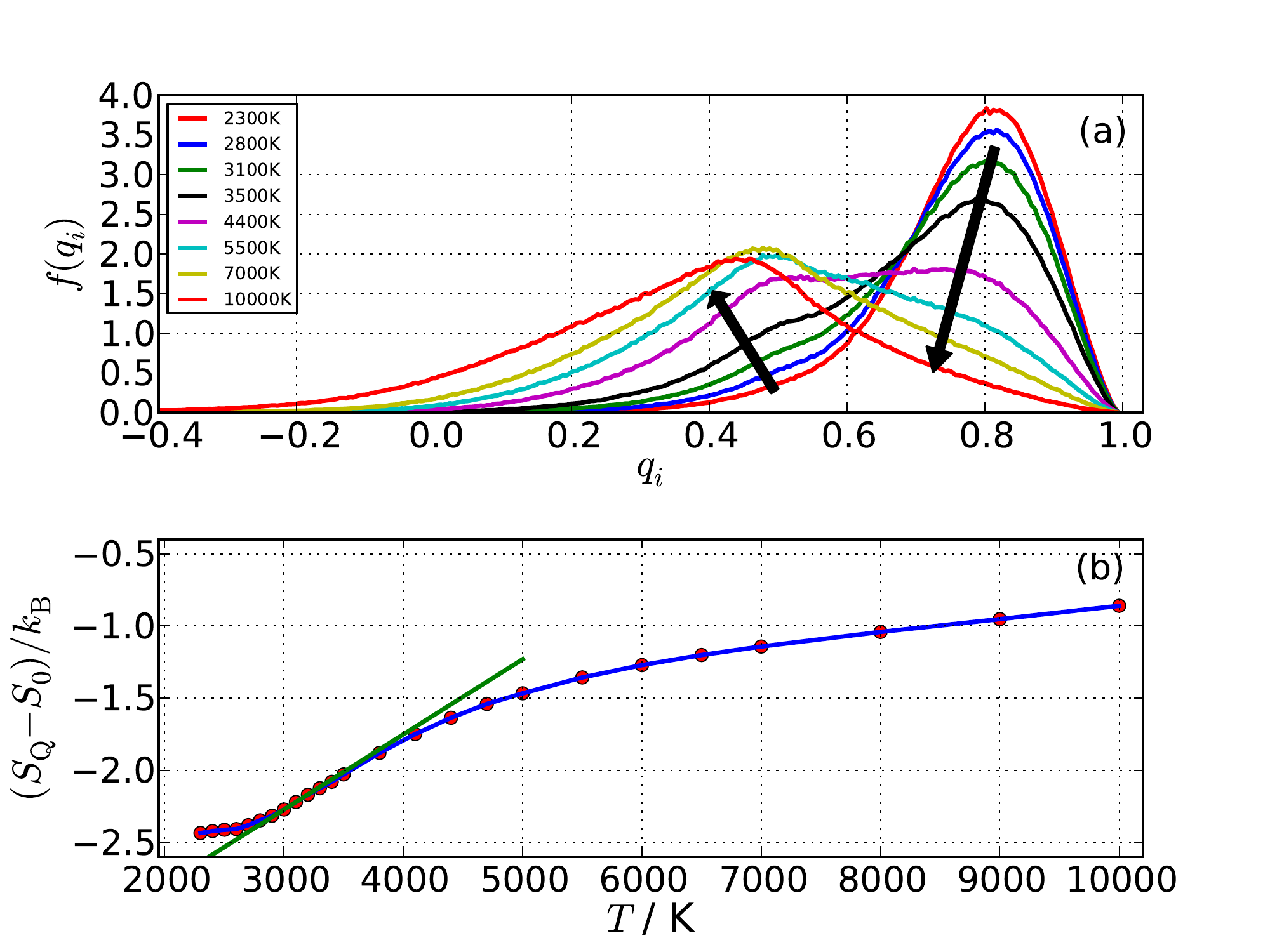}
\caption{\label{fig:tetraDistribution} (a) Distribution of the tetrahedral order parameter for various temperatures. The arrows mark the evolution of the distribution with increasing temperature. (b) Tetrahedral entropy difference ($(S_Q-S_0)/k_{\mathrm{B}}$ versus temperature $T$. The solid line is the tangent at the temperature $T\approx3100$\,K, where the gradient has a maximum.}
\end{figure}

The structural changes that cause the density anomaly can be seen in the tetrahedral order parameter $q_i$ of the silicon atoms. Fig.~\ref{fig:tetraDistribution}(a) shows the distribution $f(q_i)$ at various temperatures. For low temperatures the distribution has a sharp peak at $q_i \approx 0.80$. With increasing temperature $f(q_i)$ develops a shoulder at lower $q_i$ values, which evolves into a new peak at $q_i \approx 0.43$. At 4500--5000\,K this new peak becomes the main maximum of the distribution. These findings indicate that the degree of local order changes considerably with temperature. At low temperatures the structure is basically tetrahedral, similar to supercooled water, which exhibits roughly the same value of the tetrahedral order parameter.\cite{klameth2013} At higher temperatures the local structure is significantly less ordered. These results confirm and improve previous results by Shell \textit{et al.}\cite{shell2002} who simulated a smaller system comprised of 450 atoms over a narrower temperature range and worked at constant density rather than at constant pressure. 

The structural changes are equally well visible in the tetrahedral entropy $S_{Q}(T)$. The change in tetrahedral entropy $\frac{S_{Q}-S_{0}}{k_{\mathrm{B}}}$ is shown in Fig.~\ref{fig:tetraDistribution}(b). We see that the tetrahedral entropy strongly increases in the range 3000--5000\,K and has a turning point at $T\approx 3100$\,K, while the temperature dependence is much weaker at both lower and higher temperatures. The observed growth of $\frac{S_{Q}-S_{0}}{k_{\mathrm{B}}}$ in the intermediate temperature range indicates the breakup of the tetrahedral network upon heating. Interestingly, the temperature range of this loss of tetrahedral order agrees well with that of the density increase, providing strong evidence that these structural changes are closely related to the density anomaly. 

\subsection{Dynamics}

\begin{figure}[ht]
\centering
\includegraphics[scale=0.4]{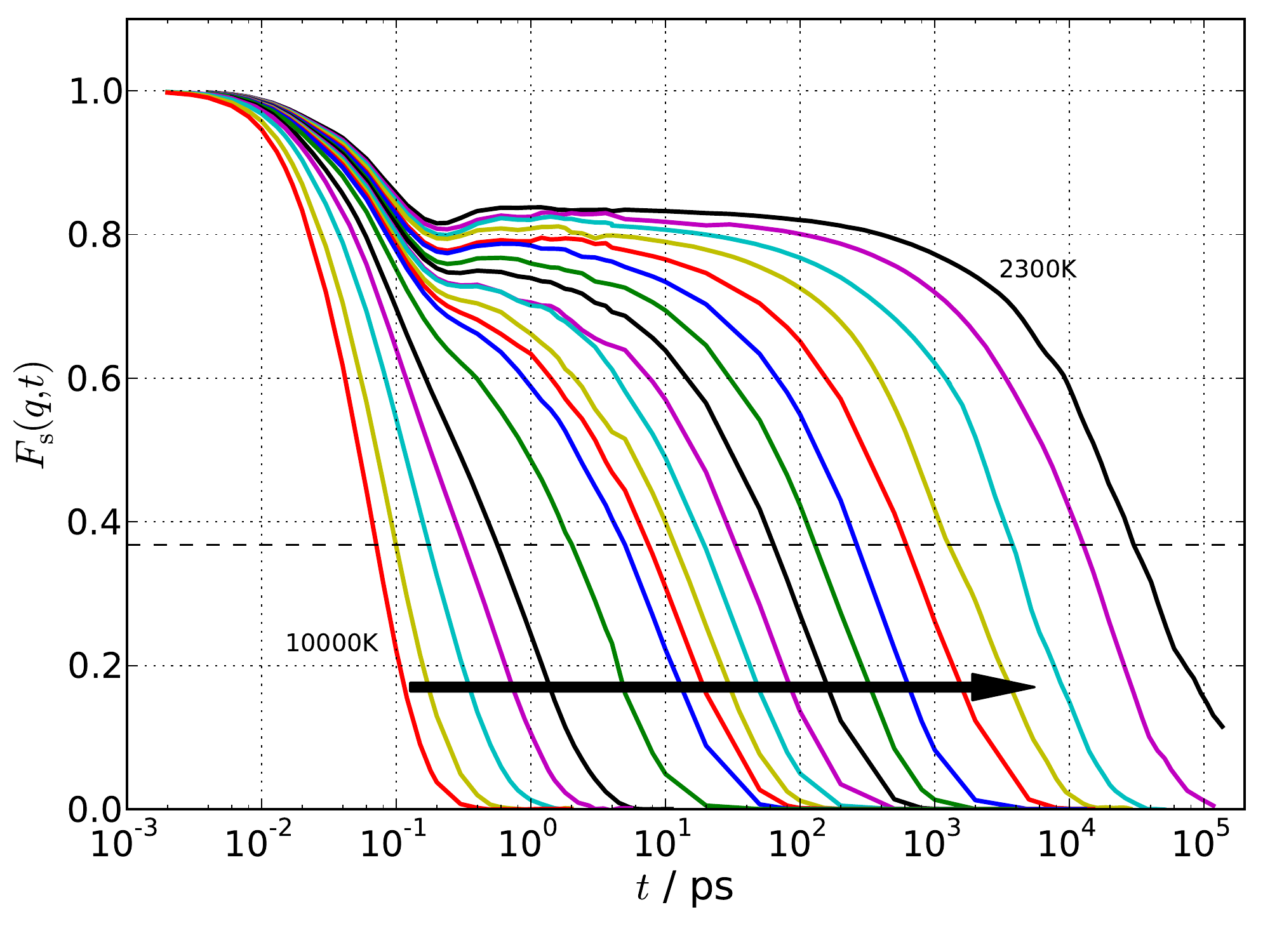}
\caption{\label{fig:isf} ISF $F_s(q,t)$ for the silicon atoms for several temperatures. The horizontal dashed line marks a value of $\frac{1}{e}$, as used for the determination of correlation times from $F_s(q,\tau)=\frac{1}{e}$. The arrow indicates increasing temperature. 
}
\end{figure}

Fig.~\ref{fig:isf} displays $F_s(q,t)$ for a wide range of temperatures (2300\,K-10000\,K). The ISF is shown for the silicon atoms, but it shows qualitatively similar behavior for the oxygen atoms (not shown), consistent with findings in previous studies on BKS silica.\cite{ horbachRelaxation2001, vogel2004} We see that $F_s(q,t)$ develops the characteristic two-step signature of glass-forming liquids upon cooling. A short-time decay due to vibrational dynamics is followed by a long-time decay due to structural relaxation, which strongly slows down upon cooling resulting in an extended intermediate plateau regime at sufficiently low temperatures. At the crossover between the vibrational and plateau regimes, i.e, at ca. $10^{-1}-10^0\,$ps, the ISF shows an oscillatory behavior, which was related to the boson peak by some scientists.\cite{horbachHigh2001}

\begin{figure}[ht]
\centering
\includegraphics[scale=0.4]{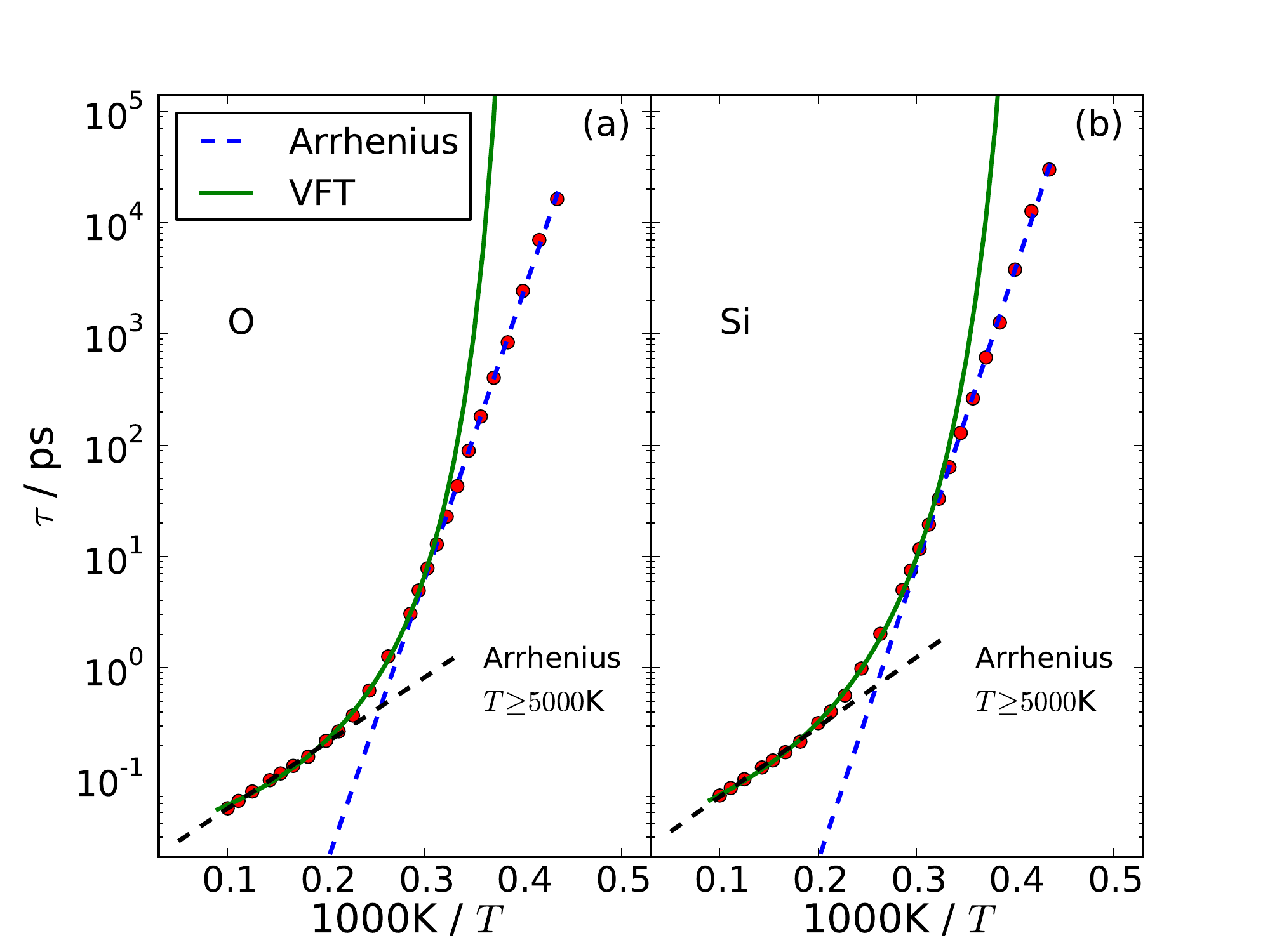}
\caption{\label{fig:tau}  Correlation times $\tau$ over inverse temperature for (a) oxygen atoms and (b) silicon atoms, together with fitted VFT  (solid) and Arrhenius  (dashed) laws.
}
\end{figure}

Fig.~\ref{fig:tau} shows the correlation times $\tau$ of the silicon and oxygen atoms as a function of inverse temperature. For both types of atoms, we observe a crossover in the temperature dependence at $T_{\mathrm{cross}}= 3200-3300$\,K. While the data are well described by a VFT fit above this crossover, they follow an Arrhenius law at lower temperatures. Table~\ref{tab:FitValues} shows the fit parameter. The VFT fits suggest a divergence of the correlation times near $T_\infty\approx2180\,$K for silicon and $T_\infty\approx2340\,$K for oxygen atoms. Alternatively, a power-law divergence predicted by the mode-coupling theory was used to describe the slowdown of the dynamics in the high-temperature regime.\cite{horbachRelaxation2001, vogel2004} In any case MD simulations consistently show that BKS silica cannot be considered as a strong glass-former at $T>T_\mathrm{cross}$. Our Arrhenius fits at $T<T_\mathrm{cross}$ yield activation energies $E_{\mathrm{A}}$ of 5.3 and 5.1\,eV for the silicon and oxygen atoms, respectively. These values compare reasonably well with the activation energies from previous experimental\cite{brebec1980,mikkelsen1984} and computational\cite{vogel2004,horbach1999,horbachRelaxation2001, heuer2004} approaches. 
Interestingly, at temperatures $T\geq 5000$\,K  correlation times obey again an Arrhenius law (see dashed line in Fig.~\ref{fig:tau}). The activation energies $E_{\mathrm{A}}$ in this regime are 1.2\,eV, which is far lower than in the supercooled regime. 

\begin{table}[ht]
\centering
\begin{tabular}{c|c|c|c|c|c|}
 & $T_{\mathrm{cross}}$ & $E_{\mathrm{A}}$ & $E_{\mathrm{VFT}}$ & $T_{\infty}$ & $E_{\mathrm{A}}$ ($T\geq 5000$\,K) \\ \hline
O & 3300K & 5.1\,eV & 0.47\,eV & 2340\,K & 1.2\,eV\\
Si & 3200K & 5.3\,eV & 0.57\,eV & 2180\,K & 1.2\,eV\\
\end{tabular}
\caption{\label{tab:FitValues} Fitting values for Arrhenius and VFT functions in Fig.~\ref{fig:tau}}
\end{table}

The observed crossover from VFT to Arrhenius behavior at $T_{\mathrm{cross}}= 3200-3300$\,K was also reported in previous MD work on BKS silica.\cite{horbach1999, horbachRelaxation2001, vogelLetter2004, vogel2004, saika2001, saikaFree2004, heuer2004} It yields evidence for the existence of a FS transition of this model. Previous studies rationalized this effect by a crossover between two liquid phases of silica\cite{saika2001} or an existence of a cutoff of the potential energy landscape.\cite{heuer2004} Our simulations cover for the first time extended dynamic and temperature ranges that include clear Arrhenius behavior at lower temperatures and clear VFT behavior at higher temperatures.

\begin{figure}[ht]
\centering
\includegraphics[scale=0.4]{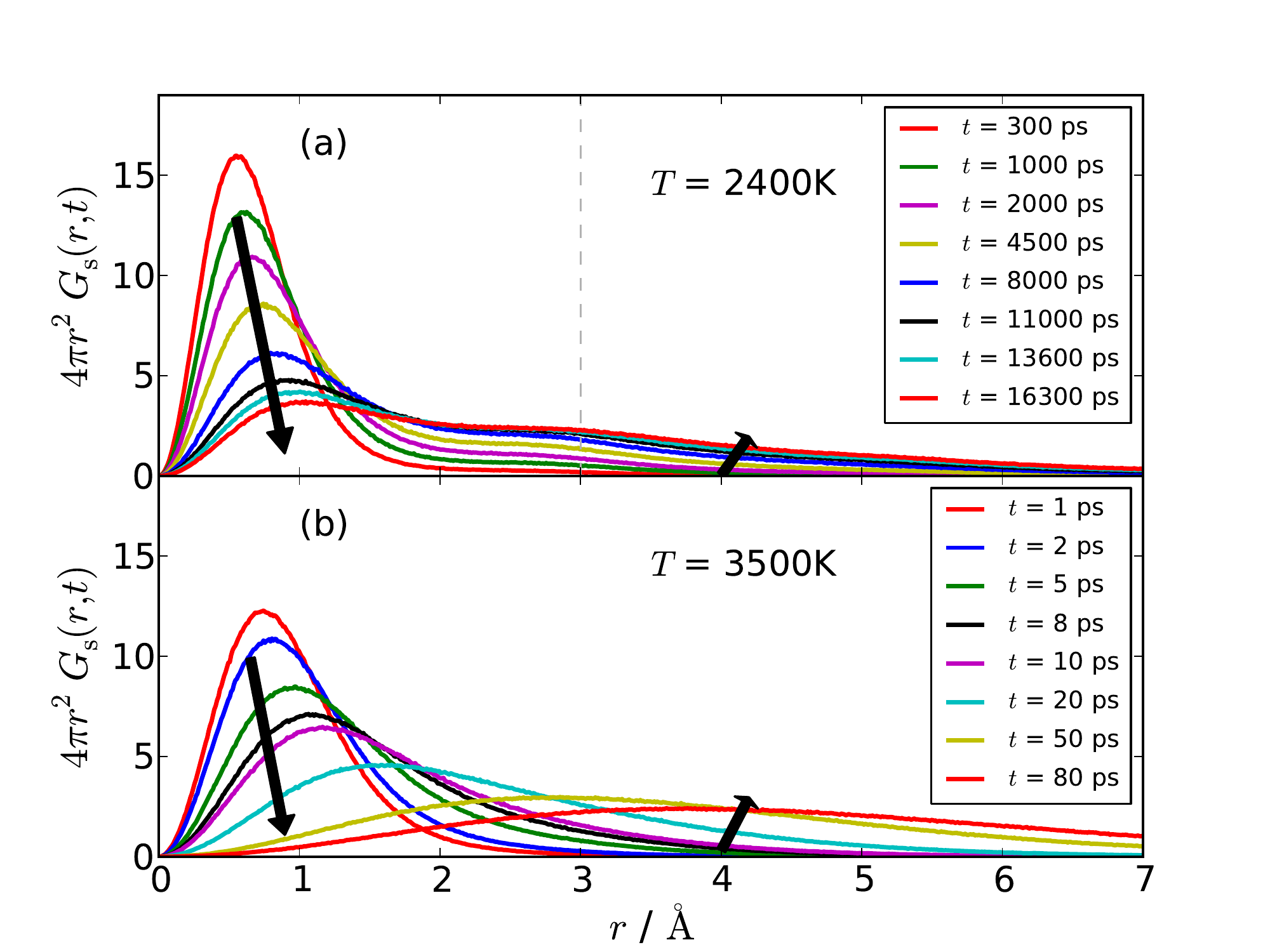}
\caption{\label{fig:gselfDist} Van Hove correlation function of oxygen atoms for various time intervals $t$ at (a) $T=2400$\,K and (b) $T=3500$\,K. At both temperatures, the considered time intervals are of the same order of magnitude as the correlation time $\tau$. In panel (a), the vertical dashed line marks the position of a second maximum at ca. 3.0\,\AA. \ 
}
\end{figure}

To investigate the mechanism for the structural relaxation, we analyze the self part of the van Hove correlation function. Fig.~\ref{fig:gselfDist} shows the time evolution of $4\pi r^2G_\mathrm{s}(r,t)$ for oxygen atoms at $2400$\,K and 3500\,K. For short time intervals $t$, the distribution is dominated by a sharp peak at around 1\,\AA, which is due to vibrational motion within local cages formed by neighbouring atoms. When extending the time interval this peak broadens. While a single-peak signature is retained at $3500$\,K, a secondary maximum grows at the expense of the primary maximum at $2400$\,K. This secondary maximum is located at $r\approx 3$\,\AA, corresponding to the oxygen-oxygen distance. These observations indicate that a hopping motion of the oxygen atoms sets in when the temperature is decreased through the crossover region at $T_{\mathrm{cross}}= 3200-3300$\,K. In contrast we do not observe hopping motion for silicon atoms. These results agree with the literature.\cite{horbach1999} The fact that only oxygen atoms show hopping motion can be understood by realizing that oxygen atoms have less mass, are twice as numerous, and interact only with half as many neighbors compared to silicon atoms in the tetrahedral structure.

\begin{figure}[ht]
\centering
\includegraphics[scale=0.4]{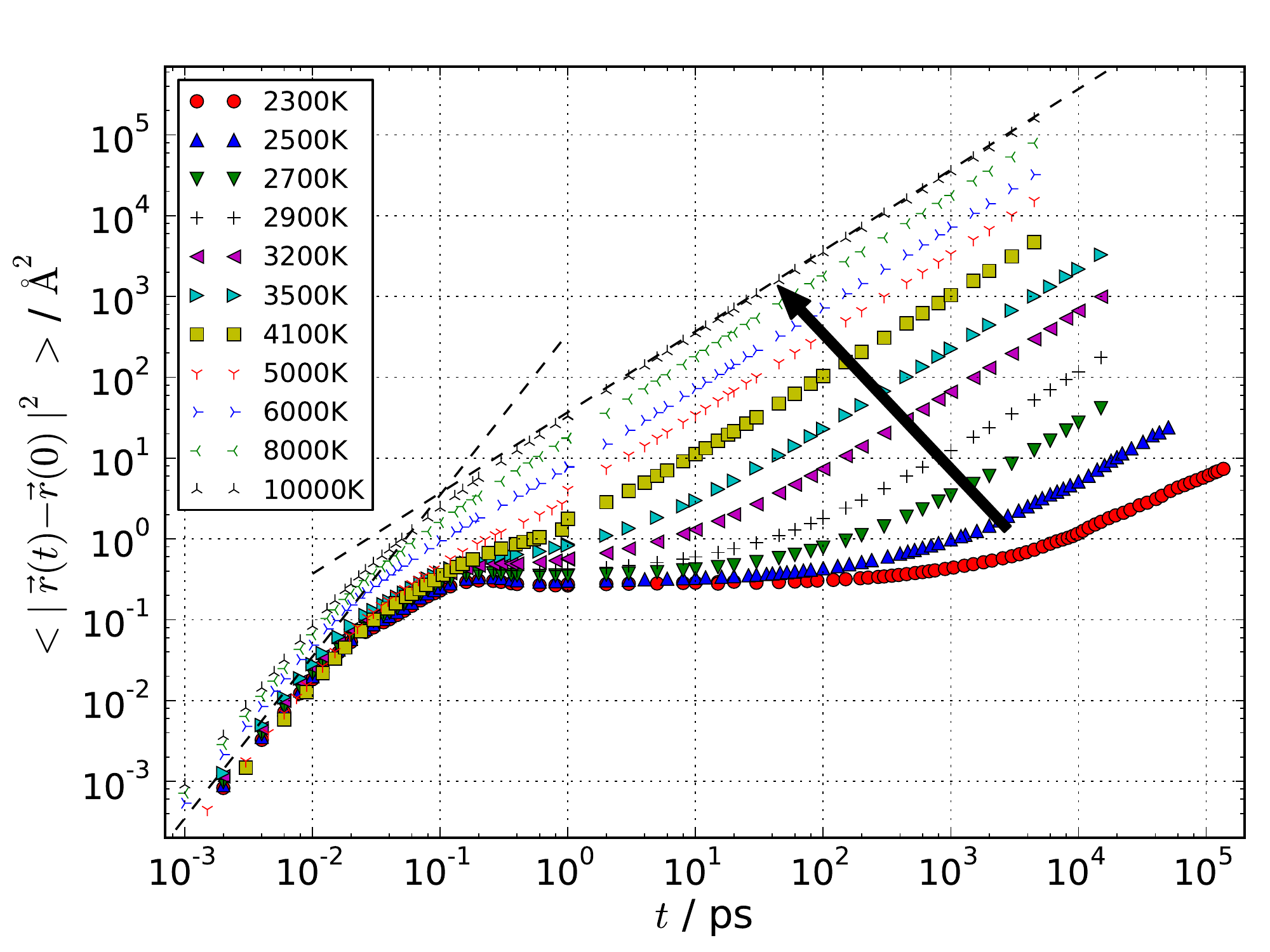}
\caption{\label{fig:vanHove} Silicon atoms mean squared displacement versus time at various temperatures. The ballistic motion with a slope of 2 for smaller times and the diffusive motion with slope $1$ for larger times can be distinguished, as indicated by the dashed lines. The arrow indicates increasing temperature.}
\end{figure}

While the van Hove correlation function gives information about the type of motion within the liquid, the average displacement helps to identify the regimes of ballistic and diffusive motion, see Fig.~\ref{fig:vanHove}, where the mean square displacement (MSD) for silicon atoms is shown. At very short time scales below 0.1\,ps, the curves have a slope of 2 due to ballistic motion. At longer times diffusive motion sets in. Between these two regimes a plateau can be seen for temperatures below ca. 3200\,K. The plateau value can be interpreted as the maximum squared distance an atom can travel within its cage. Motion over larger distances requires the rearrangement of several surrounding atoms.

\begin{figure}[ht]
\centering
\includegraphics[scale=0.4]{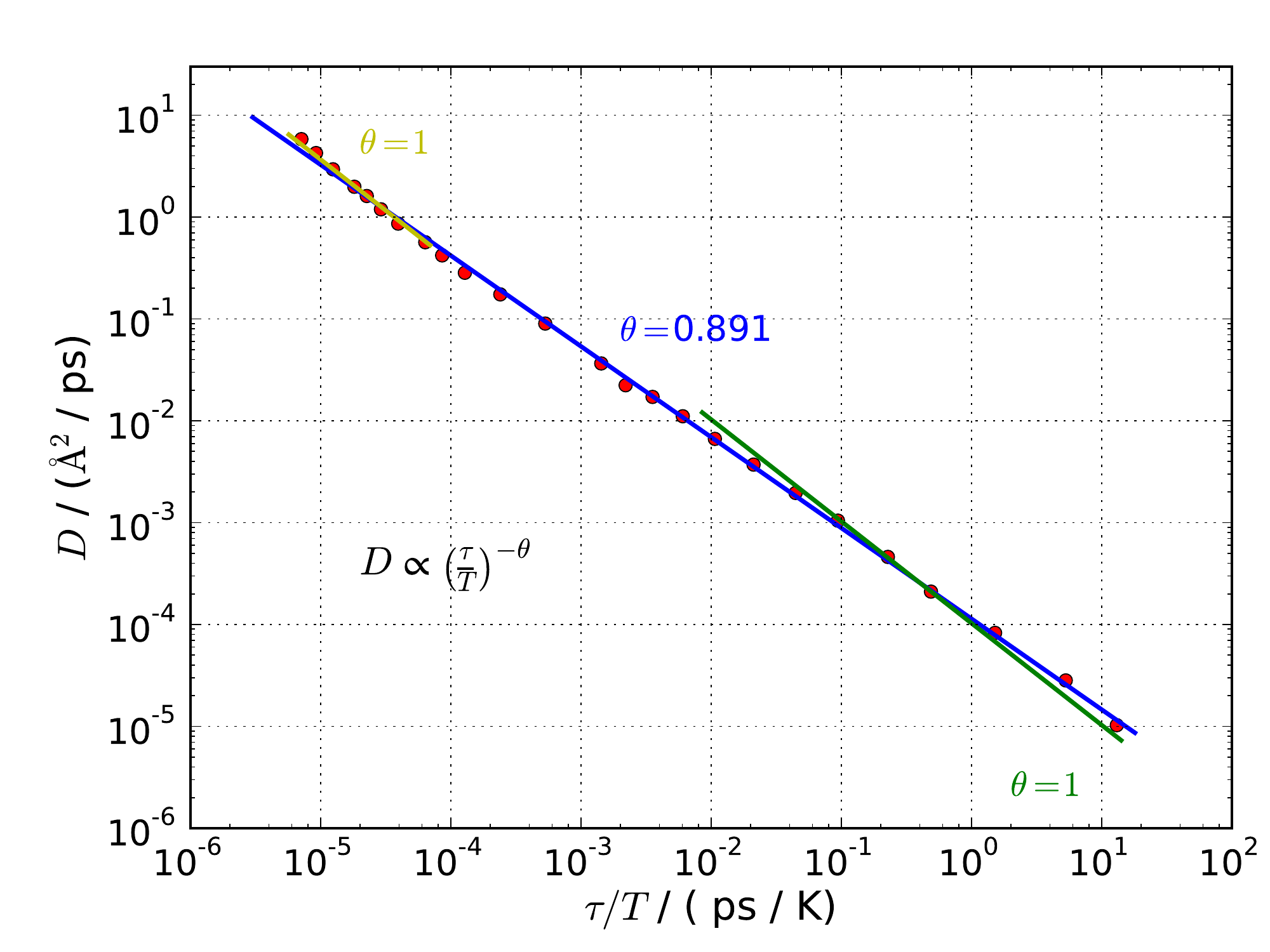}
\caption{\label{fig:Diff_vs_Tau} Diffusion coefficient versus $\frac{\tau}{T}$. The exponent $\theta \neq 1$ indicates a mild breakdown of the Stokes-Einstein relation.
The universal blue line is a fit to Eq.~(\ref{eq:stokes-einstein}) over all data points.
The green and yellow solid lines have a slope of -1, which equals $\theta=1$. Hence the Stokes-Einstein equation holds for low (green) and high (yellow) temperatures, but not in the whole temperature regime.
}
\end{figure}

Based on the data displayed in Fig.~\ref{fig:vanHove}, we calculated the diffusion coefficient by fitting the expression $6Dt$ to the MSD at sufficiently long times. The resulting diffusion coefficient $D$ is plotted as a function of the correlation time $\tau$, divided by the temperature, in Fig.~\ref{fig:Diff_vs_Tau}. Over a wide range of correlation times the data are reasonably well described by $D \propto \left(\frac{\tau}{T}\right)^{-\theta}$ with $\theta=0.89$ rather than $\theta=1$. The corresponding temperature range is 2300\,K--10000\,K. We note that this breakdown of the classical Stokes-Einstein relation in BKS silica was not observed in previous studies covering a narrower $T$ range.\cite{henritzi2015} We did not observe this breakdown either when we used a $NVT$ instead of a $NpT$ ensemble. 
Closer inspection of Fig.~\ref{fig:Diff_vs_Tau} reveals, however, that the Stokes-Einstein relation is recovered at low as well as high temperatures, where the solid lines with a slope of $\theta=1$ fit the data well. This means that the  breakdown of the Stokes-Einstein relation occurs in the same temperature regime as the non-Arrhenius behavior.

\section{Conclusion}\label{sec:Conclusion}
We performed a comprehensive MD study of the structural and dynamical features of liquid silica and found coherent evidence of a transition between two qualitatively different liquids in the vicinity of 3300\,K. The structural features show an anomalous behavior of the density, which increases with temperature between 3000\,K and 5000\,K, with the maximum slope near 3400\,K. Around this temperature, the tetrahedral order parameter changes from a high value ($Q=0.8$) to a considerably lower value ($Q=0.4$). The temperature at which the tetrahedral entropy changes fastest is around 3100\,K.  These structural changes are accompanied by dynamical changes. The correlation time shows a qualitative change from an Arrhenius behavior, i.e., from a strong glass at lower temperatures (below 3300\,K) to a Vogel-Fulcher-Tammann behavior, i.e., a fragile glass at higher temperatures. At very high temperatures ($T\geq 5000$\,K), dynamics shows again Arrhenius behavior, which means that silica becomes a simple liquid at sufficiently high temperatures. In the temperature range where the Arrhenius behavior breaks down, the Stokes-Einstein relation breaks also down, and the liquid shows anomalous diffusion with an exponent $\theta \simeq 0.9$ for the mean-square displacement. 

All our results were obtained with a pressure of 1 bar. We also performed simulations at higher pressures and found that the transition from strong to fragile behavior moves to lower temperatures. For a pressure of 100\,kbar the liquid is fragile for $T\geq 2700$\,K. This change is expected since the phase with higher density is thermodynamically favored at higher pressures. 

Taken together, our results extend considerably earlier evidence for a fragile-to-strong (FS) transition in silica. Such a transition was observed in experiments\cite{roessler1998,hess1996,sonneville2013} and simulations.\cite{saikaFree2004,vogel2004,saika2001}  \textit{Saika-Voivod et al.} found a FS transition in the potential energy surface\cite{saikaFree2004} and in the energy landscape.\cite{saika2001} \textit{Vogel et al.} discovered a FS transition in the transport coefficient.\cite{vogel2004}  \textit{Horbach et al.} \cite{horbachRelaxation2001} evaluated the correlation time, but could only see the onset of the transition from VFT to Arrhenius behavior, due to the much narrower range of temperature and time scales. The breakdown of the Stokes-Einstein relation in silica has not been reported before. The transition back to simple dynamical behavior at very high temperatures has not been reported either.

The finding of a  FS transition in simulations of small systems can be interpreted in several ways. The different possible thermodynamic scenarios compatible with the obeserved anomalies were discussed in \cite{debenedetti2003} for liquid water. \textit{Heckmann et al.} showed that these scenarios are expected for any liquid with the same structure as water,\cite{heckmann2012,heckmann2013} and which of the possible scenarios is realized depends on the values of the parameters that characterize the interactions between the molecules, and on the entropy associated with the different possible nearest-neighbor configurations. First, there can be a true thermodynamic phase transition between two liquids with different density and different degree of order. In a small system, this phase transition would be visible as a smooth crossover between the two phases, and the transition temperature can be expected to be in the vicinity of the temperature where the changes are fastest, i.e. around 3300\,K. Second, the system may have a true phase transition, but the investigated range of temperature and pressure values is beyond the critical end point of the transition line. The line where changes are fastest is the so-called Widom line, and it is the extension of the phase transition line beyond the critical point. Third, the critical point may be at zero temperature, so that the system shows no true phase transition, but it shows the anomalies caused by this critical point.  Our data are compatible with any of these three situations. Future simulation studies or experiments will hopefully be capable of settling this interesting question whether silica or other network-forming liquids show a true thermodynamics phase transition in the liquid phase. If the observed FS transition is a true phase transition in the thermodynamic limit, this transition should become sharper with increasing system size. In contrast to water, silica has the advantage that it can be studied in the supercooled phase in experiments because it does not crystallize that easily. The reason is that the ion bonds in silica have a higher energy than the hydrogen bonds in water, making the barriers higher that have to be overcome when forming the crystal. For the experimental studies this means in particular that no nanoconfinement is required in order to prevent crystallization.

\section{Acknowledgments}
This work was performed within the DFG research unit FOR 1583 and supported by grant numbers Dr300/11-2 and Vo-905/9-2.

\end{document}